\documentclass[aps,showkeys,twocolumn,showpacs,amsmath,amssymb]{revtex4}
\usepackage{color}
\usepackage{graphicx}

\input{Qcircuit}
\begin{document}
\title{One way quantum bit commitment through classical correlation}
\pacs{ 03.67.Dd, 03.67.Ac, 03.67.-a}
\keywords{Entanglement, Bit-commitment, Correlation and Cryptography.}
\author{Sriram Prasath E.}
\email{sridif@gmail.com} \affiliation{Maulana Azad National Institute of Technology, Bhopal - 462051, India}

\author{Prasanta K. Panigrahi}
\email{prasanta@prl.res.in}
\affiliation{Indian Institute of Science Education and Research (IISER) Kolkata, Mohanpur, BCKV Campus Main Office, Mohanpur - 741252, India}
\affiliation{Physical Research
Laboratory, Navrangpura, Ahmedabad - 380 009, India}

\begin{abstract}
  A one way partial quantum bit commitment protocol is developed, using states with built-in classical correlation, completely independent of entanglement. It involves concealing information in a set of mutually non-orthogonal states and revealing it through measurements on a set of product states that are mutually orthogonal. Given $2^N$ choices to commit from, the protocol encodes each choice in a $N+1$ qubit state, from $2^N$ non-orthogonal states. A previously agreed upon $N$ qubit state corresponding to each choice, when coupled with the $N+1$ qubit state, yields an element belonging to a set of orthogonal product states, which can be deterministically distinguished. The protocol is demonstrated by implementing it for a `coin-toss' game. A procedure to enhance security of the protocol is explicated, increasing the number of qubits required. Thus a modification is suggested to reduce this required number.

\end{abstract}
\maketitle
 Bit commitment is a basic operation in cryptography, to which many practically important cryptographic tasks, such as quantum coin tossing \cite{coin}, quantum gambling \cite{gamb}, quantum oblivious mutual identification \cite{id}, quantum oblivious transfer \cite{trans} and two-party secure computations \cite{compute}, can be reduced. In a bit commitment process between two parties, Alice and Bob, Alice first commits her bit with some evidence to Bob. At the time of revealing, Alice publishes her commitment and Bob verifies it with the evidence. A bit commitment is perfect, if it is both binding and concealing. It is binding, if Alice cannot change her committed bit, once she has made the decision. It is concealing, if Bob cannot eavesdrop on the committed bit before the legal revealing phase. Classically, bit commitment can be achieved with computational security, but not with information theoretic security \cite{wies}. It was shown by Bennett and Brassard \cite{bennet} that, quantum bit commitment can be defeated by the use of entangled states. Attempts were made to construct secure bit commitment protocols \cite{ brassard2}, which were again proven to be insecure, independently by Mayers \cite{mayer} and Lo and Chau \cite{lochau}. It was shown that, an entangled attack, akin to that of \cite{bennet} defeats all bit commitment protocols \cite{ariano}. Bit commitment using at most one ebit of entanglement has been presented \cite{David}. Measurement based semi-quantum key distribution protocols have been presented  \cite{michel1, michel2, srik1}, in which one of the parties is classical, but the protocols are robust against an eavesdropping attempt. A protocol to hide certain quantum states, in non-orthogonal states has been explicated \cite{Ahmed}. Proof that there exists a protocol, whose security increases exponentially  with the number of systems, which is locally non-classical and lacking entanglement, has been recently given \cite{how}.\\

 Entanglement is a non classical feature, involving non-trivial quantum correlation in a multipartite quantum system. Apart from entanglement, quantum mechanics is endowed with tensor product spaces in Hilbert space. In the product space, one can generate non-trivial classical correlation through orthogonality, where the subsystems are non-orthogonal \cite{bennet3, niset}. The present protocol is developed, using states with built-in classical correlation, completely independent of entanglement. It involves associating commitment choices to a set of mutually non-orthogonal states for concealing the information. It is revealed through measurements on a set of product states that are mutually orthogonal. Given $2^N$ choices to commit from, the protocol encodes each choice in a $N+1$ qubit state, from $2^N$ non-orthogonal states. Previously agreed upon $N$ qubit states corresponding to the choices, when coupled with the specific $N+1$ qubit states, yield elements belonging to a set of orthogonal product states, which can be deterministically distinguished, thus revealing the commitment choice. \\

There are two parties Alice and Bob, who wish to have a quantum bit commitment protocol to communicate between themselves. For a given number of commitment choices, $2^N $, one starts with $2N+1$ qubits. Alice has $N+1$ qubits and Bob has $N$ qubits. Alice subsequently makes a set $B_i$, with $1\leq i\leq 2^N$, where each set contains $2^N$ elements.  As is evident, choosing one state from a set containing $2^N$ choices increases security. Alice and Bob have an initial agreement regarding the commitment choice and the corresponding set $B_i$. Depending on her choice, Alice chooses one member from the appropriate set and sends it to Bob. Since, there are equal number of sets and commitment choices, each choice can be associated with one particular set. This association is initially agreed upon; the contents of these sets are known to both the parties. Each of these states is a superposition of two terms, such that all the elements are orthogonal. Each of these sets, in principle can have $2^{N+1}$ orthogonal states, however only $2^N$ states, with equal superposition amplitudes, are retained. Terms in the superposition have been chosen such that, any element of a particular set, has two non-orthogonal elements in each of the other sets, ensuring that there are no suitable measurement basis, which can reveal the information about the chosen state with certainty. In principle, this overlap region maybe enlarged with larger number of terms in the superposition. In the initial agreement, corresponding to each set $B_i$, a specific N qubit state is associated by Bob. This is required to distinguish the overlapping sets, in a bigger Hilbert space of mutually orthogonal states. At this stage, Bob communicates classically his commitment, after which Alice reveals her initial commitment.\\

In case of disagreement, Bob can generate a $N$ qubit state, in reference to their initial agreement, to form a $2^{2N+1}$ product space $2^{N}$ mutually orthogonal states, for distinguishing the initial state with certainty. The set of $ N$ qubit state generated by Bob has the stabilizer codes $(\sigma_X)^{\otimes N}$ and $(M)^{\otimes N}$, where $M \in (\sigma_Z,I)$ and the weight is even, i.e., number of non-identity operations is even. The circuit for generation of such states is shown in Fig. 1. The qubits in the circuit are denoted by $|i\rangle_1, |j\rangle_2, |k\rangle_3, |l\rangle_4....,|m\rangle_N \in (|0\rangle ,|1\rangle)$, corresponding to $2^N$ commitment choices, so that each of these can be associated with a commitment choice. A schematic representation of the protocol is shown in Fig. \ref{eca0}. The initial $N+1$ qubit state sent by Alice contains two terms as also the $N$ qubit state generated by Bob, thus the final product state contains four terms, described by the stabilizers $I^{\otimes 3}\otimes(\sigma_X)^{\otimes N}$ and $I^{\otimes 3}\otimes(M)^{\otimes N}$, where $M \in (\sigma_Z,I)$ and the weight is even.\\

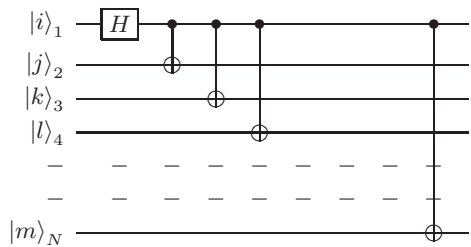
\begin{figure}[h]
\label{fig:CircuitDiagram}
	\leavevmode
\centering	
\Qcircuit @C=1em @R=.7em {
\lstick{\ket{i}_1}&\gate{H}&\ctrl{1}&\ctrl{2}&\ctrl{3}&\qw &\qw &\qw &\ctrl{6}&\qw\\
\lstick{\ket{j}_2}&\qw     &\targ   &\qw &\qw   &\qw &\qw &\qw &\qw&\qw\\
\lstick{\ket{k}_3}&\qw     &\qw     &\targ&\qw  &\qw &\qw &\qw &\qw&\qw\\
\lstick{\ket{l}_4}&\qw     &\qw     &\qw  &\targ&\qw &\qw &\qw &\qw&\qw\\
\lstick{-}&\push{-}&\push{-}&\push{-}&\push{-}&\push{-}&\push{-}&\push{-}&\push{-}\\
\lstick{-}&\push{-}&\push{-}&\push{-}&\push{-}&\push{-}&\push{-}&\push{-}&\push{-}\\
\lstick{\ket{m}_N}&\qw     &\qw    &\qw  &\qw  &\qw &\qw &\qw &\targ&\qw\\
}
	\caption{Circuit to generate N qubit state in the revealing phase.}
	
\end{figure}

\begin{figure*}
%\begin{center}
\includegraphics[scale=0.4]{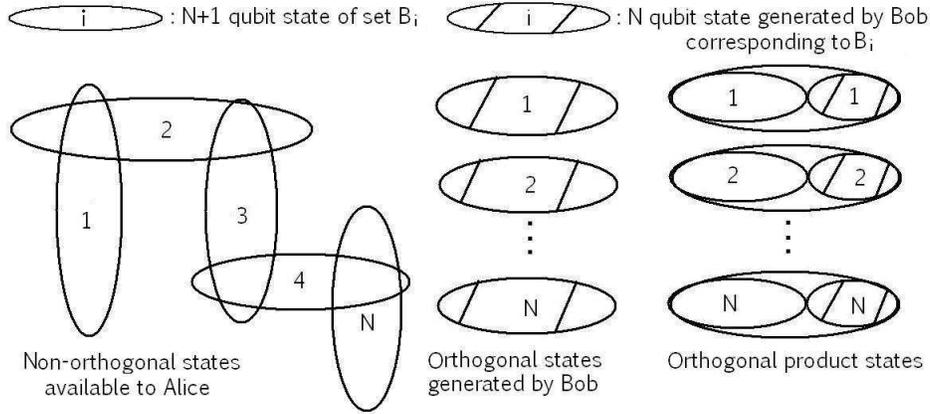}
\caption{\label{eca0} Schematic representation of the protocol.}
%\end{center}
\end{figure*}

\begin{figure*}
%\begin{center}
\includegraphics[scale=0.45]{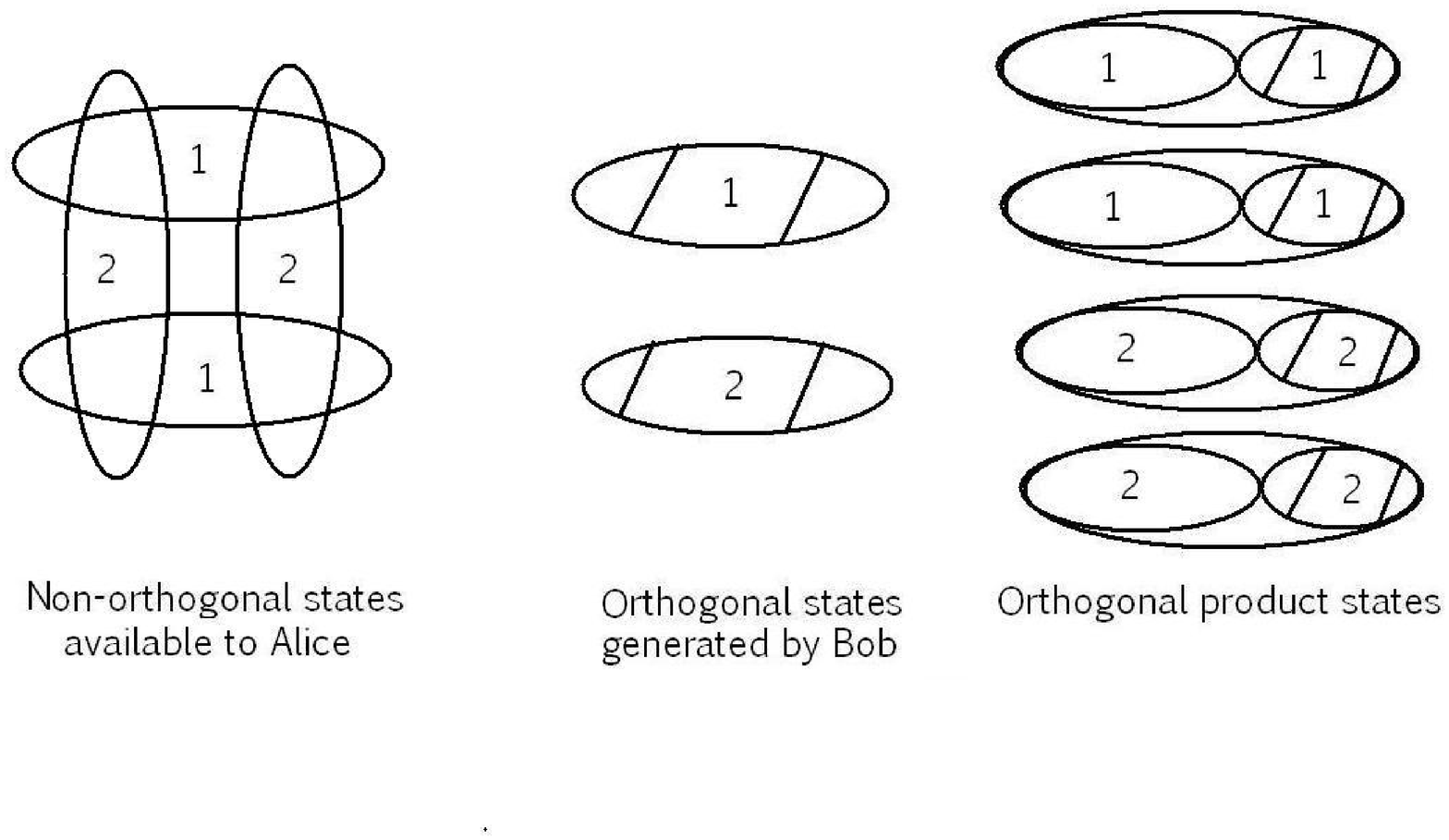}

\caption{\label{eca} Two party `coin-toss' game.}
%\end{center}
\end{figure*}

We now explicate the protocol for a `coin-toss' game with two possible commitment choices. It is a two player game, played by Alice and Bob. Alice has to toss a coin and Bob has to predict the outcome. But as they are separated in space, it makes a bit commitment protocol necessary for fair play. In our protocol, we use a two qubit state at Alice's end and another one qubit state at Bob's end to form a three qubit product state. Since there are only two possible outcomes in a coin toss, the number of choices is $ 2$ and  $N=1$. In the commitment phase, Alice sends a two qubit state to Bob, depending upon the output of her coin toss results, based on the table I. In case Alice tosses the coin and finds head, she can choose to send either $\frac{|01\rangle + |00\rangle}{\sqrt{2}}$ or $\frac{|10\rangle + |11\rangle}{\sqrt{2}}$, both of which are elements of $B_1$. Then they move on to the second phase; through a classical channel Bob guesses the outcome of the coin toss. Alice then reveals the result, thus indirectly revealing the set $B_1$. In case of mistrust they move on to the third phase; Bob generates a one qubit state corresponding to the value of $i$ revealed by Alice and couples this state with the two qubit state already in his possession to ascertain Alice's commitment. It can be readily seen that all the four product states, shown in table II are mutually orthogonal. Thus Bob can make a three qubit measurement in this basis and verify Alice's commitment. A schematic representation, as per the previous conventions, is shown in Fig. \ref{eca}.\\
\begin{table}[h]
\caption{\label{tab1}The initial agreement table for the commitment phase.}
\begin{tabular}{|c|c|c|}
\hline
\textbf{The information to be sent}&\textbf{Set} &\textbf{Elements of $B_i$}\\
\hline

head   &$B_1$& $\frac{|01\rangle + |00\rangle}{\sqrt{2}}$\\
        &   & $\frac{|10\rangle + |11\rangle}{\sqrt{2}}$\\
\hline
tail   &$B_2$& $\frac{|01\rangle + |10\rangle}{\sqrt{2}}$\\
        &   & $\frac{|00\rangle + |11\rangle}{\sqrt{2}}$\\

\hline

\end{tabular}
\end{table}

\begin{table}[h]
\caption{\label{tab2}The initial agreement for the revealing phase.}
\begin{tabular}{|c|c|c|}
\hline
\textbf{$N+1$ qubit state}& \textbf{N qubit state}&\textbf{product state}\\
\hline
$\frac{|01\rangle + |00\rangle}{\sqrt{2}}$& $\frac{|0\rangle + |1\rangle}{\sqrt{2}}$& $\frac{|010\rangle +|000\rangle +|011\rangle +|001\rangle}{2} $\\
$\frac{|10\rangle + |11\rangle}{\sqrt{2}}$& $\frac{|0\rangle + |1\rangle}{\sqrt{2}}$& $\frac{|100\rangle +|110\rangle +|101\rangle +|111\rangle}{2} $\\
\hline
$\frac{|01\rangle + |10\rangle}{\sqrt{2}}$& $\frac{|0\rangle - |1\rangle}{\sqrt{2}}$& $\frac{|010\rangle +|000\rangle -|011\rangle -|001\rangle}{2} $\\
$\frac{|00\rangle + |11\rangle}{\sqrt{2}}$& $\frac{|0\rangle - |1\rangle}{\sqrt{2}}$& $\frac{|000\rangle +|110\rangle -|001\rangle -|111\rangle}{2} $\\
\hline
\end{tabular}
\end{table}

 The protocol developed here is compliant with several desirable requirements. Two sets of states $B_1$ and $B_2$ are available corresponding to the commitment choices. Each element in $B_1$ has two non-orthogonal states in $B_2$, likewise for each element in $B_2$. Since there are only two states in each of the sets, every element in one set is mutually non-orthogonal to every other element in another set. Hence a suitable three qubit measurement basis at Bob's end, can determine the parent set of the 2 qubit state, with complete fidelity. The protocol also ensures that before the revealing phase, Bob has no way to form product states and use alternate basis, to extract information with complete certainty. If Bob intends to form the the three qubit state before the legal revealing phase, the plausible final states should not be distinguishable. For illustration, Bob couples the initial two qubit state with an inappropriate $N$ qubit states, i.e., not in accordance with the initial agreement, as shown in table \ref{tab3}. The product states formed by appropriate $ N $ qubit states as shown in table II and inappropriate $N$ qubit states as shown in table III are mutually non-orthogonal. From the product states so formed, it can be inferred that, the fidelity with which Bob can find the parent set $B_i$, is $ \frac{1}{2}$. This is equivalent to the probability of an arbitrary guess, implying no information has been found, before the legal revealing phase. Such inappropriate coupling made by Bob, will also occur when Alice cheats. These non-orthogonal states, can be misused by Alice, to deceive Bob with $\frac{1}{2}$ fidelity. This factor can be reduced, if Alice is made to split the information into many blocks, and each block is communicated through a distinct $N+1$ qubit state, as per the protocol. If the number of blocks is $K$, then this fidelity is $({\frac{1}{2}})^K$. The protocol, thus becomes more secure with the increase in number of blocks. It can be verified that, fidelity does not depend on the value of $N$, but only on the value of $K$, implying that a smaller $N$ and a larger $K$ values, ensures a better security.\\

\begin{table}[h]
\caption{\label{tab3} States that can be formed apart from the ones in revealing phase.}

\begin{tabular}{|c|c|c|}
\hline
\textbf{$N+1$ qubit}& \textbf{N qubit state}&\textbf{product state}\\
\hline
$\frac{|01\rangle + |00\rangle}{\sqrt{2}}$& $\frac{|0\rangle - |1\rangle}{\sqrt{2}}$& $\frac{|010\rangle +|000\rangle -|011\rangle -|001\rangle}{2} $\\
$\frac{|10\rangle + |11\rangle}{\sqrt{2}}$& $\frac{|0\rangle - |1\rangle}{\sqrt{2}}$& $\frac{|100\rangle +|110\rangle -|101\rangle -|111\rangle}{2} $\\
\hline
$\frac{|01\rangle + |10\rangle}{\sqrt{2}}$& $\frac{|0\rangle + |1\rangle}{\sqrt{2}}$& $\frac{|010\rangle +|000\rangle +|011\rangle +|001\rangle}{2} $\\
$\frac{|00\rangle + |11\rangle}{\sqrt{2}}$& $\frac{|0\rangle + |1\rangle}{\sqrt{2}}$& $\frac{|000\rangle +|110\rangle +|001\rangle +|111\rangle}{2} $\\
\hline
\end{tabular}

\end{table}

 When there is a need to send many blocks of information, the above protocol can be modified to reduce the number of qubits required. A set $S$ containing $N+1$ qubit states in their computational basis as their elements, can be considered. In total there are $2^{N+1}$ such elements are possible, out of which $2^N$ are initially related to each of the commitment choices. Alice can select either set $S$ or sets $B_i$ as a parent set, which will be used for encoding the commitment. It is clear that all the element of the set $S$ are orthogonal, thus information appears unconcealed. But with the lack of knowledge of the Alice's selection, set S or sets $B_i$, Bob has no way to determine the parent set, since every element of set S will have two non-orthogonal elements in each of the sets $B_i$. Alice has to additionally reveal the parent set with the commitment. Bob has to prepare the $N$ qubit state if sets $B_i$ are chosen for commitment, in case of set S, he can measure the $N+1$ qubit state in the computational basis. Thus the number of required qubits reduces to $N+1$ qubits for $2^N$ choices. The protocol's security inversely depends upon the probability of $S$ being the parent set, as Bob upon guessing the parent set correctly as $S$, has the complete information.\\

   In conclusion, a protocol for one way quantum bit commitment, is presented for a general situation, involving two parties, who initially share a quantum channel between them. The protocol is developed using states with built-in classical correlation. Since it is independent of entanglement, it cannot be subjected to entanglement based attack. It involves concealing information in a set of mutually non-orthogonal states and revealing it through measurements on a set of product states that are mutually orthogonal. In order to demonstrate the protocol, we have implemented it in a `coin-toss' game, played between two players separated in space.
   A procedure to increase the security is given, ensuring Alice does not use the non orthogonal product states formed to her advantage. This procedure, eventually increases the number of qubits required, but an alternate protocol, with a modification is suggested to reduce this number. In future, we would like to develop such protocols for secure function evaluation and other cryptographic tasks which can be reduced to quantum bit commitment.\\

   The authors would like to acknowledge Dr. R. Srikanth for discussions. SP acknowledges the warm hospitality of the Indian Institute of Science and Research, Kolkata, India where the work was concluded and Open Course Ware (OCW) for making available a comprehensive set of study materials.


\begin{thebibliography}{0}
\bibitem{coin} G. Brassard and C. CrŽepeau, Advances in Cryptology: Proceedings of Crypto 90, Lecture Notes in Computer Science vol. 537 (Springer Verlag, Berlin, 1991), p. 49.
\bibitem{gamb} L. Goldenberg, L. Vaidman, and S. Wiesner, Phys. Rev. Lett. {\bf82}, 3356 (1999).
\bibitem{id} C. CrŽepeau and L. Salvail, in \textit{ Advance in Cryptology- Proceedings of Eurocrypt} (Springer Verlag, Berlin,(1995)) p. 133.
\bibitem{trans} C. CrŽepeau, J. Mod. Optics {\bf41}, 2445 (1994).
\bibitem{compute} H.-K. Lo, Phys. Rev. {\bf56}, 1154 (1997).
\bibitem{wies} S. Wiesner, SIGACT News 15, 78 (1983).
\bibitem{bennet}C. H. Bennett and G. Brassard, in \textit{ Proceedings of IEEE International Conference on Computers, Systems, and Signal Processing, Bangalore, India} (IEEE press, (1985)), p. 175179.
\bibitem{brassard2}G. Brassard, C. C\'{r}epeau, R. Jozsa, and D. Langlois, in \textit{Proceedings of the 34th Annual IEEE symposium on foundations of Computer Science}, (Los Alamitos,(1993)), p. 362.
\bibitem{mayer}D. Mayers, Phys. Rev. Lett. {\bf78}, 3414 (1997).
\bibitem{David} D. P. DiVincenzo, D. W. Leung, and B. M. Terhal, in \textit{IEEE Transaction on Information Theory}, (vol.48 No.3 (2002)), p. 580-598.
\bibitem{lochau}H. K. Lo and H. F. Chau, Phys. Rev. Lett. {\bf78}, 3410 (1997).

\bibitem{ariano}G. M. D'Ariano, D. Kretschmann, D. Schlingemann, and R. Werner, Phys. Rev. A. {\bf76}, 032328 (2007).

\bibitem{michel1}M. Boyer, D. Kenigsberg, and T. Mor, Phys. Rev. Lett. {\bf99}, 140501 (2007).
\bibitem{michel2}M. Boyer, R. Gelles, D. Kenigsberg, and T. Mor, Phys. Rev. A. {\bf79}, 032341 (2009).
\bibitem{srik1} R. Srikanth, Physica Scripta {\bf 70}, 343 (2004).
\bibitem{Ahmed} A. Younes, arXiv:quant-ph/0907.3532v1  (2008).

\bibitem{how} H. Barnum, O. C. O. Dahlsten, M. Leifer, and B. Toner, \textit{Information Theory Workshop}, (2008) p. 386-390.
\bibitem{bennet3} C. H. Bennett, D. P. DiVincenzo, C. A. Fuchs, T. Mor, E. Rains, P. W. Shor, J. A. Smolin, and W. K. Wootters, Phys. Rev. A {\bf59}, 1070 (1999).
\bibitem{niset} J. Niset and N. J. Cerf, Phys. Rev. A {\bf74}, 052103 (2006).

\end{thebibliography}
\end{document}